\documentclass[pra,twocolumn,floatix,amssymb,showpacs,amsmath,superscriptaddress]{revtex4-1}
\usepackage{siunitx}

\usepackage{graphicx}
\usepackage{epsfig}
\usepackage{amsmath,amssymb}
\usepackage{ dsfont }
\usepackage{color}
\usepackage[utf8]{inputenc}
\usepackage[T1]{fontenc}
\usepackage[english]{babel} 
\usepackage{float}
\usepackage{times}
\usepackage{dsfont}
\usepackage{hyperref,amsmath,amssymb,amscd}
\usepackage{amssymb,amsmath,amsfonts,amsthm}
\usepackage{mathtools}
\usepackage{verbatim}
\usepackage{enumerate}
\usepackage{graphicx}
\graphicspath{{figs/}}
\usepackage{hyperref}
\usepackage{bbm}
\usepackage{booktabs}

\usepackage[caption=false]{subfig}

\usepackage{color}
\definecolor{dred}{rgb}{.8,0.2,.2}
\definecolor{ddred}{rgb}{.8,0.5,.5}
\definecolor{dblue}{rgb}{.2,0.2,.8}
\definecolor{dgreen}{rgb}{.2,0.5,.2}

\newcommand{\be}{\begin{equation}}
\newcommand{\ee}{\end{equation}}

\newcommand{\bea}{\begin{eqnarray}}
\newcommand{\eea}{\end{eqnarray}}

\begin{document}

\newcommand{\quadr}[1]{\ensuremath{{\not}{#1}}}
\newcommand{\quadrd}[0]{\ensuremath{{\not}{\partial}}}
\newcommand{\slpar}{\partial\!\!\!/}
\newcommand{\gtrescero}{\gamma_{(3)}^0}
\newcommand{\gtresuno}{\gamma_{(3)}^1}
\newcommand{\gtresi}{\gamma_{(3)}^i}

\title{Arbitrary function resonance tuner of the optical microcavity with sub-MHz resolution}

\author{Xu-Sheng Xu}
\address{State Key Laboratory of Low-Dimensional Quantum Physics, Department of Physics, Tsinghua University, Beijing 100084, China}

\author{Hao Zhang}
\address{State Key Laboratory of Low-Dimensional Quantum Physics, Department of Physics, Tsinghua University, Beijing 100084, China}

\author{Min Wang}
\address{State Key Laboratory of Low-Dimensional Quantum Physics, Department of Physics, Tsinghua University, Beijing 100084, China}

\author{Dong Ruan}
\address{State Key Laboratory of Low-Dimensional Quantum Physics, Department of Physics, Tsinghua University, Beijing 100084, China}

\author{Gui-Lu Long}
\email{Correspondence and requests for materials should be addressed to G.L.L.:gllong@tsinghua.edu.cn}

\address{State Key Laboratory of Low-Dimensional Quantum Physics, Department of Physics, Tsinghua University, Beijing 100084, China}

\address{Beijing Information Science and Technology National Research Center, Beijing 100084, China}

\address{Beijing Academy of Quantum Information Sciences, Beijing 100193, China}

\date{\today}

\begin{abstract}
The resonance frequency of an optical whispering gallery mode (WGM) microcavity is extremely important in its various applications. Many efforts have been made to fine tune this parameter. Here, we report the design and implementation of a function resonance tuner of an optical microcavity with resolution about 650 kHz (7 pm @ 1450 nm band), 20\% of the optical WGM linewidth. A piezoelectric nano-positioner is used to mechanically compress the microsphere in its axial direction. The ultrafine frequency tuning is achieved benefitting from the much less changes in the axial direction than equatorial semiaxes of the microsphere and the sub-nanometer resolution of the nano-positioner. The tuning of the resonance can be made to an arbitrary function, dynamically, with near perfect accuracy. We have demonstrated the periodically tuning of resonance in the sine and sigmoid function respectively, both with over 99\% fitting accuracy. This work expands the application of microresonators greatly, especially microspheres with ultrahigh quality factor, in multi-mode coupling system or time-floquet system.
\end{abstract}

\maketitle
\section{Introduction}
Whispering gallery mode (WGM) microcavities, with ultrahigh quality factor (Q) and small mode volume \cite{vahala_mc}, have shown tremendous applications in modern physics, such as optical sensing \cite{yl_sensing, xyf_sensing, sqh_sensing, xyf_sensing2, yl_sensing2}, cavity QED \cite{mk_qed, vahala_qed, hailin_qed}, nonlinear optics \cite{bf_nonlinear, xyf_nonlinear} and optical information processing \cite{hailin_opto, kipp_opto, wm_opto2, vahala_opto, wm_opto, cy_ln, or_ln, dch_opto}. Silica WGM microcavities, such as microspheres \cite{hailin_deformed, frank_biosensing, wang_scibut}, microrings \cite{wc_microring} and microbubbles \cite{sumetsky_mb, yy_mb}, are good platforms to generate strong light-matter interaction. Among them, the microsphere has higher quality factor and it is easier to be fabricated. However, the optical resonance frequency is fixed by the fabrication process, which is not favourable in applications. Actually, it is vital to tune the microcavity resonance frequency to match between multiple coupling systems, such as chiral atom-light interaction \cite{lxf_chiral} and electromagnetically induced transparency (EIT) experiments \cite{yl_eit, fxd_eit}. Due to the ultrahigh quality factor of WGM, matching the resonance frequencies with ultrafine resolution is also required. In other experiments like time-floquet modulation \cite{ktt_tf}, it is a key issue to tune the system parameters periodically. Owing to all of these requirements, periodically tuning the optical resonance of microspheres with ultrahigh resolution becomes an essential technique.

Significant methods have been proposed to match resonance frequencies and periodically tune the frequency. For example, the temperature tuning method has been used to couple the nitrogen vacancy centers with a microsphere \cite{hailin_tempe} and to observe the EIT-like phenomenon in two or more coupled WGM resonators system \cite{yl_eit, wt_eit}, which is an easy and effective approach to adjust the resonance. Aerostatic pressure could also tune the frequency of microcavities, especially microbubble \cite{oliver_gas2, oliver_gas}. By adjusting the air pressure in microbubble, the geometry and the refractive index can be modified, and the optical modes will be redshifted or blueshifted. However, the tuning resolution of these methods are not high enough for experiments with ultrahigh Q microcavities and the heating or cooling of microcavities is a process with relatively long relaxation time, which is not suitable for real-time tuning experiments. Mechanical strain tuning is another technique to shift the resonance frequency of microcavities \cite{hailin_mec, dch_mec, wolf_mec, har_mec}. A deformed double-stemmed microsphere was used in these experiments, and the resonance frequency was tuned by stretching the stems with piezo-driven nano-positioner. This method can effectively tune the resonance over a free spectral range, but the tuning of optical WGM resonance frequency below \SI{1}{\MHz} has not been reported.

\begin{figure}[htbp]
\centering
\includegraphics[width=\linewidth ]{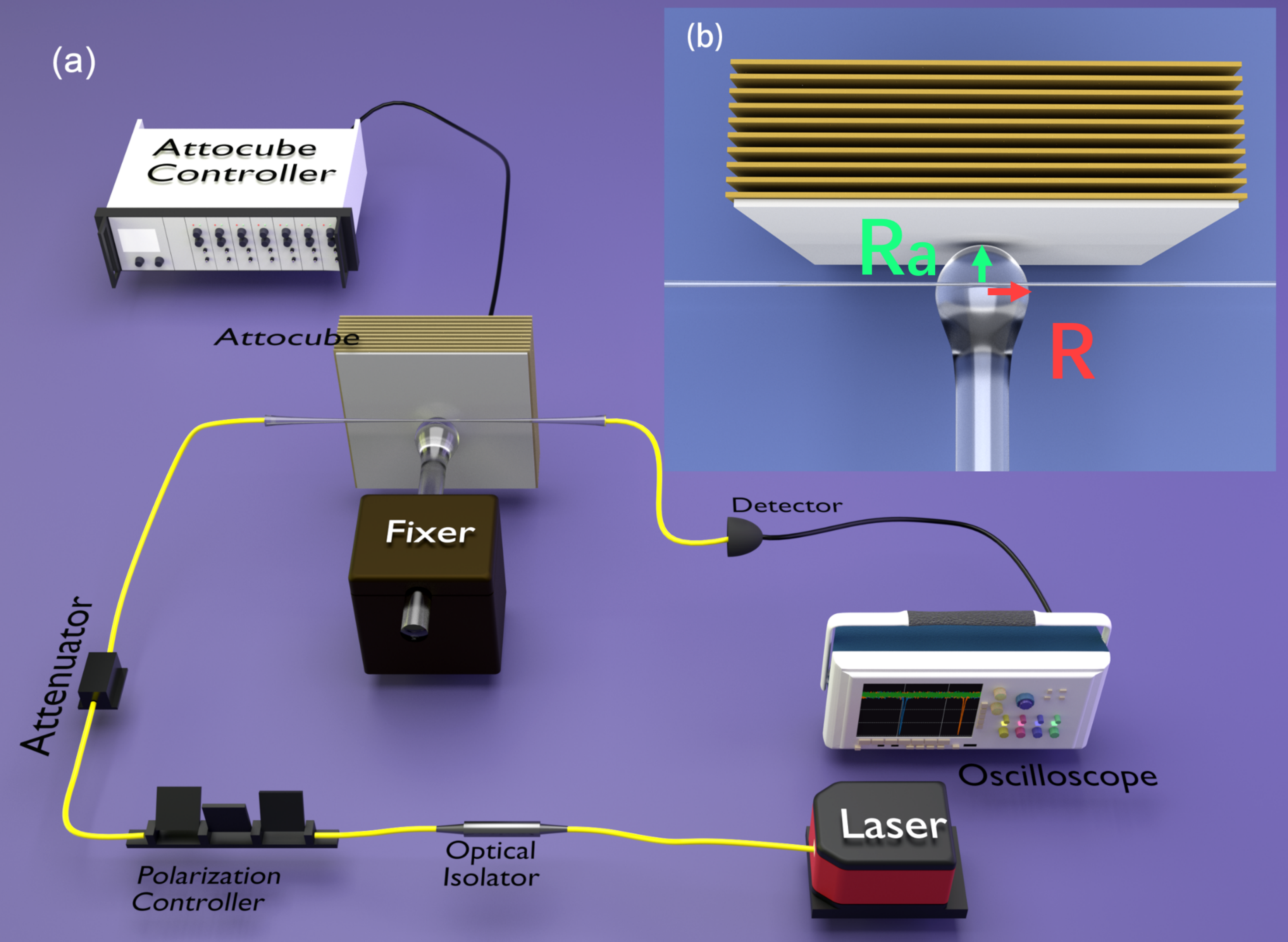}
\caption{(a) Schematic of our AFRETOM system. A CO$_2$ laser reflowed microsphere with radius $R=$ \SI{103.50}{\um} was fixed on a stage and a \SI{1450}{\nm} tunable laser was coupled in and out of the sphere with a tapered fiber. To tune the resonance of the microsphere, a piezoelectric nano-positioner was used to compress the microsphere in the axial direction. (b) Top view of the compression setup.}
\label{fig:setup}
\end{figure}

In this paper, we established an arbitrary function resonance tuner of the microsphere (AFRETOM) with kHz resolution based on mechanical strain tuning. Meanwhile, the nonlinear tuning of the resonance frequency as a sine wave or a sigmoid function with ultrahigh precision was first reported. Microspheres fused from single mode fiber, with quality factor around $0.44\times 10^8$ to $0.65\times 10^8$ were used in our experiment, and a piezoelectric nano-positioner was applied to compress the microsphere in the axial direction. Due to the high precision of the nano-positioner with sub-nanometer movement per step, we tuned the resonance frequency with resolution about \SI{650}{\kHz} in \SI{1450}{\nm} wave band, \SI{20}{\percent} of the WGM linewidth, i.e. \SI{7}{\pm} in wavelength change. We also demonstrated that the tuning of frequency depends linearly on the displacement of the nano-positioner, and we successfully shifted the resonance frequency with sine and sigmoid waves periodically with above \SI{99}{\percent} accuracy. Our work provides an alternative way for improving the experimental feasibility of studying complicated multi-modes system with ultrahigh quality factor, such as optical wavelength conversion \cite{oskar_conv, Zhang_shortcuts, dch_conv}.

\begin{figure*}
\centering
\includegraphics[width=\linewidth]{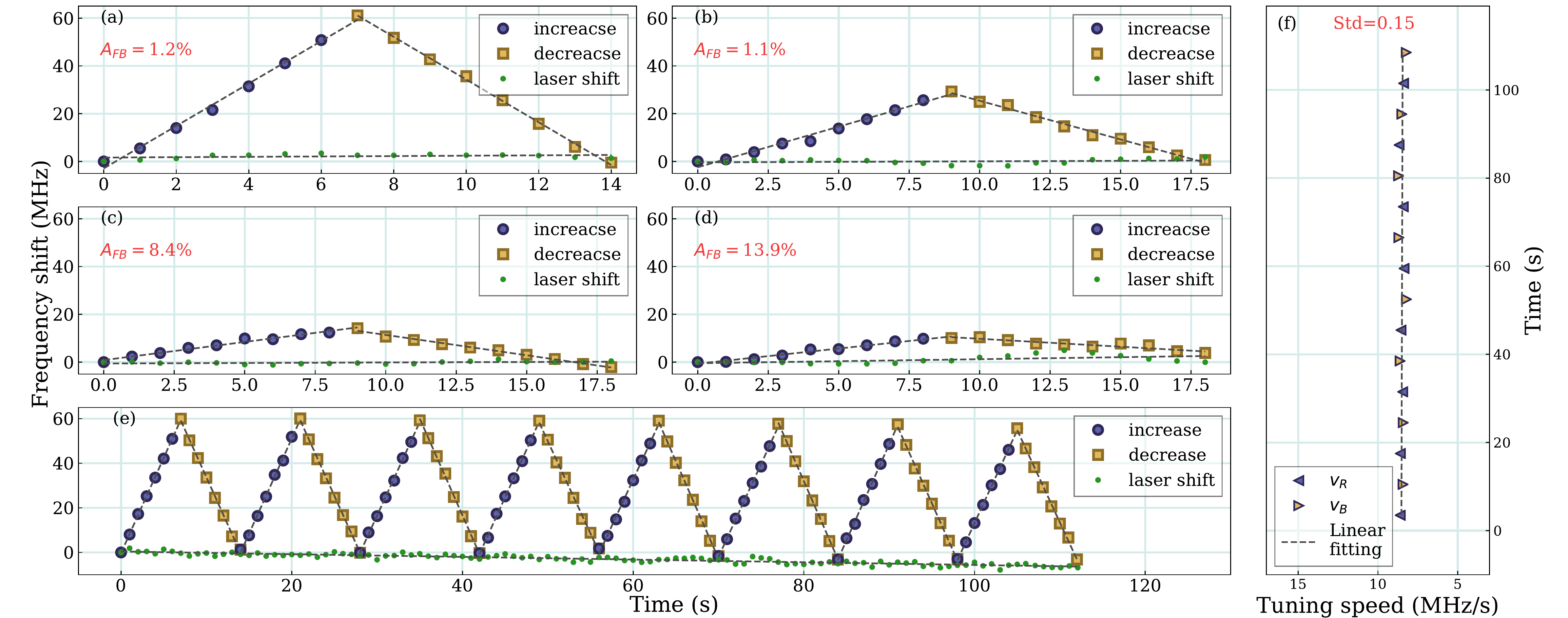}
\caption{Four different working voltage ranges $V_m$ and steps $V_s$, for (a) $V_m=$ \SI{3.5}{\V}, $V_s=$ \SI{0.5}{\V}, (b) $V_m=$ \SI{1.8}{\V}, $V_s=$ \SI{0.2}{\V}, (c) $V_m=$ \SI{0.9}{\V}, $V_s=$ \SI{0.1}{\V}, (d) $V_m=$ \SI{0.45}{\V}, $V_s=$ \SI{0.05}{\V}, where blue circle and yellow square are increasing and decreasing of the working voltage, respectively. The forward-backward asymmetry $A_{FB}$ was also given here. (e) The periodic modulation of the resonance frequency in a period of $T=$ \SI{14}{\second}. (f) The tuning speed of every period of (e). With $V_s=$ \SI{0.5}{\V}, the tuning speed $v_R$ and $v_B$ are around \SI{8.53}{\MHz/\second}, with standard deviation $\sigma=0.15$.}
\label{fig:u_d_p}
\end{figure*}

\section{Results}
For a microsphere with radius $R$, the resonance of its WGM is approximated by $\nu=mc/(2 \pi n R)$, where $m$ and $n$ are the mode number and refraction index, respectively. For a certain WGM with fixed mode number $m$, the resonance frequency could be red-tuned (or blue-tuned) by increasing (or decreasing) the radius $R$. In our experimental setup , a piezoelectric nano-positioner (model ANPZ101, Attocube) was installed in front of the microsphere with distance around nanometers , as shown in Fig. \ref{fig:setup}. The nano-positioner could move back and forth under the control of input voltage provided by the Attocube controller. Once the nano-positioner touches the microsphere, the radius of microsphere will be changed correspondingly, so will the resonance frequency. Here a tunable laser with wave length $\lambda=$ \SI{1450}{\nm} and linewidth less than \SI{200}{\kHz} was utilized as the probe light. The laser is coupled into and out of the microsphere by a single-mode optical tapered fiber with diameter around \SI{1}{\um}. The output signal is received by a photodetector and transferred to an oscilloscope. In our system, an Electro-Optic Modulator (EOM) is used to measure the resonance shift. For all of our measurements, the probe laser power is as weak as \SI{0.88}{\uW}, in order to avoid the thermal broadening of the Lorentz lineshape transmission spectrum and the thermal drift of resonant frequency caused by a high power scanning laser, which will introduce errors into the measurement of the resonance frequency shift.

The microspheres are fabricated from a single mode fiber. Since there is a strong absorption line of 10.6 um laser for silica material, one end of the silica fiber could be exposed in a low-power CO$_2$ laser, and melt under high temperature. With the surface tension, the melted silica will form a nearly perfect sphere at that end of fiber. For this kind microsphere, the quality factor is about \SI{0.63d8}, which is high enough for various experiments.

\subsection{The repeatability and precision of the frequency tuning}
We periodically tuned the optical resonance of microspheres by applying a triangular waveform to the Attocube controller, for the potential application of microspheres in time-floquet system. What's more, the repeatable tuning is also necessary for experiments that need to be done multiply times for a certain microcavity. In order to measure the asymmetry of frequency tuning during compression and release, four working voltage ranges were chosen as $V_m=$ \SI{3.5}{\V}, \SI{1.8}{\V}, \SI{0.9}{\V} and \SI{0.45}{\V} with steps $V_s=$ \SI{0.5}{\V}, \SI{0.2}{\V}, \SI{0.1}{\V} and \SI{0.05}{\V} per second, respectively. Taking Fig. {\ref{fig:u_d_p}(a)} for example, first we increased the working voltage to \SI{3.5}{\V} in \SI{7}{\second}, and the resonance frequency would be red-tuned with speed of $v_{r}$. $v_{b}$ is the blue tuning speed of the resonance when we decreased the working voltage in the next \SI{7}{\second}. We supposed that the laser shift was constant during this measurement, so we also measured the speed of laser shift $v_{l}$ just before we applied any working voltage to the nano-positioner. Considering the laser shift, we modified the tuning speeds to $v_R=v_{r}-v_{l}$ and $v_{B}=v_{b}+v_{l}$ for red and blue tuning respectively. Here we define the forward-backward asymmetry as
\begin{equation}
 A_{FB}=\frac{v_{R}-v_{B}}{v_{R}+v_{B}}=\frac{v_{r}-v_{b}-2v_{l}}{v_{r}+v_{b}}
\label{eq:asymmetry},
\end{equation} which means that we can entirely repeat the same tuning process if $A_{FB}=0$. The asymmetry in our experiment is caused by two factors. First, laser drifting with random noise, which is inevitable in our experiment. The $A_F$ will increase with smaller step $V_s$, especially when the frequency tuning of one step is comparable to the laser drifting. Second, about \SI{5}{\percent} forward/backward step asymmetry of the nano-positioner. Besides the single period behaviour, we also measured the frequency tuning for 8 periods [See Fig. \ref{fig:u_d_p}(e)]. The tuning speed $v_R$ and $v_B$ are almost constant during these 8 periods, with standard deviation $\sigma=0.15$ [See Fig. \ref{fig:u_d_p}(f)], which shows that our method is a good way to achieve high precision periodic frequency tuning.

\begin{figure}[htbp]
\centering
\includegraphics[width=\linewidth]{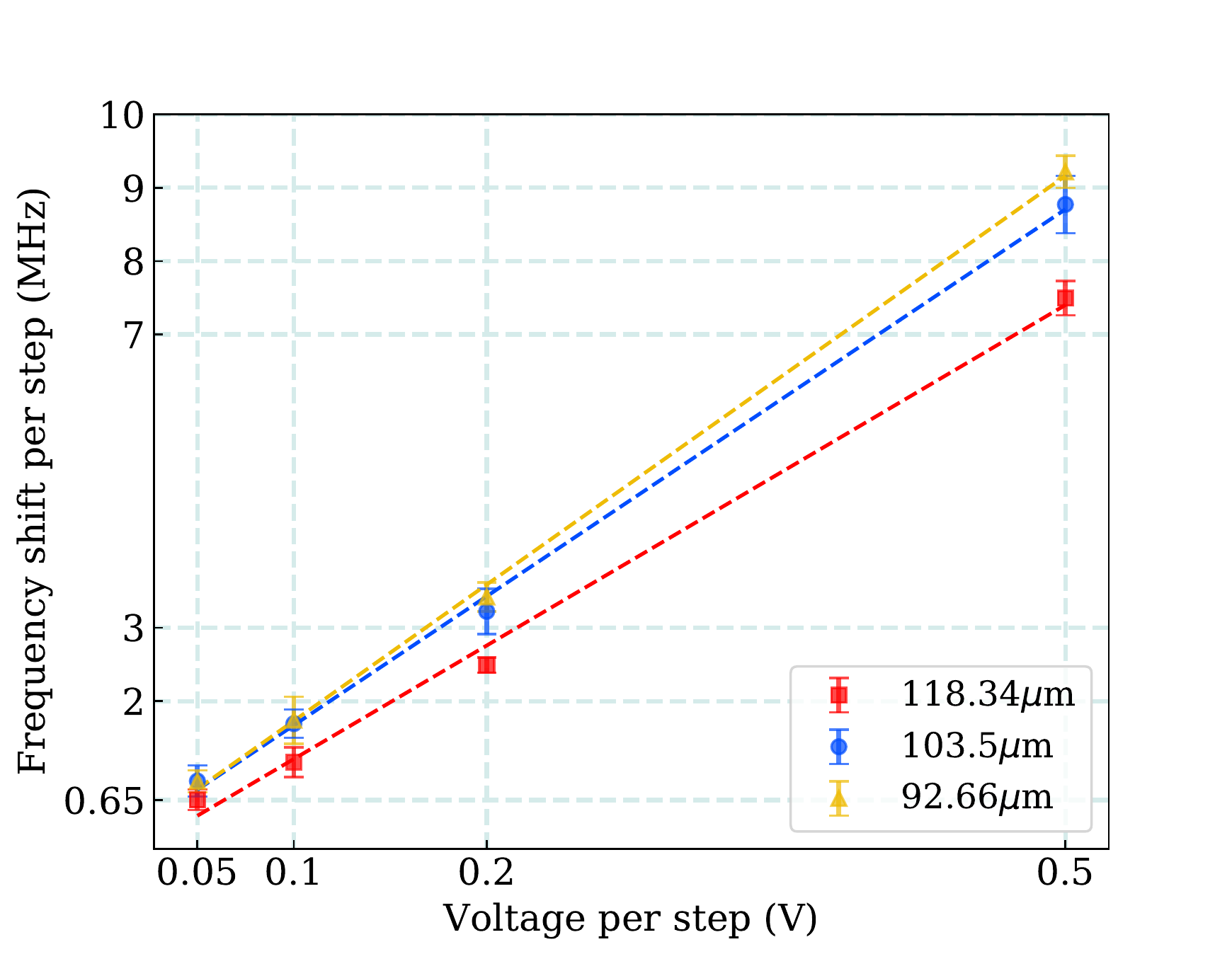}
\caption{The tuning speed of three microspheres with different size under different working voltage steps. For a bigger radius microsphere cavity, $R=$ \SI{118.34}{\um} in our experiment, we obtained an ultrahigh tuning resolution, with $\Delta \nu=$ \SI{650}{\kHz} or $\Delta \lambda =$ \SI{7}{\pm}, which is less than a half line width of the WGM in this experiment.}
\label{fig:three_ball}
\end{figure}

In order to reveal the relationship between the tuning speed and the size of the microsphere, we also studied another two spheres with $R=$ \SI{92.66}{\um} and \SI{118.34}{\um}. By applying different steps of working voltage, we can find the different frequency tuning response for each one. The experiment result [see Fig. \ref{fig:three_ball}] shows that, with the same working voltage step, the resonance frequency tuning is finer for a bigger sphere. In this work we achieved a \SI{650}{\kHz} frequency tuning for a microsphere with $R=$ \SI{118.34}{\um}, which is far more precise than previous work \cite{hailin_mec, xf_therm}.

\subsection{Maximum shift of resonance frequency}
The maximum shift of the resonance frequency and the change of quality factor are two important indicators for our method. To ensure that the performance of the microsphere does not change too much, the quality factor should be kept as stable as possible. We studied the quality factor change first, and then we found the maximum mode frequency shift for the microsphere.

\begin{figure}[htbp]
\centering
\includegraphics[width=\linewidth]{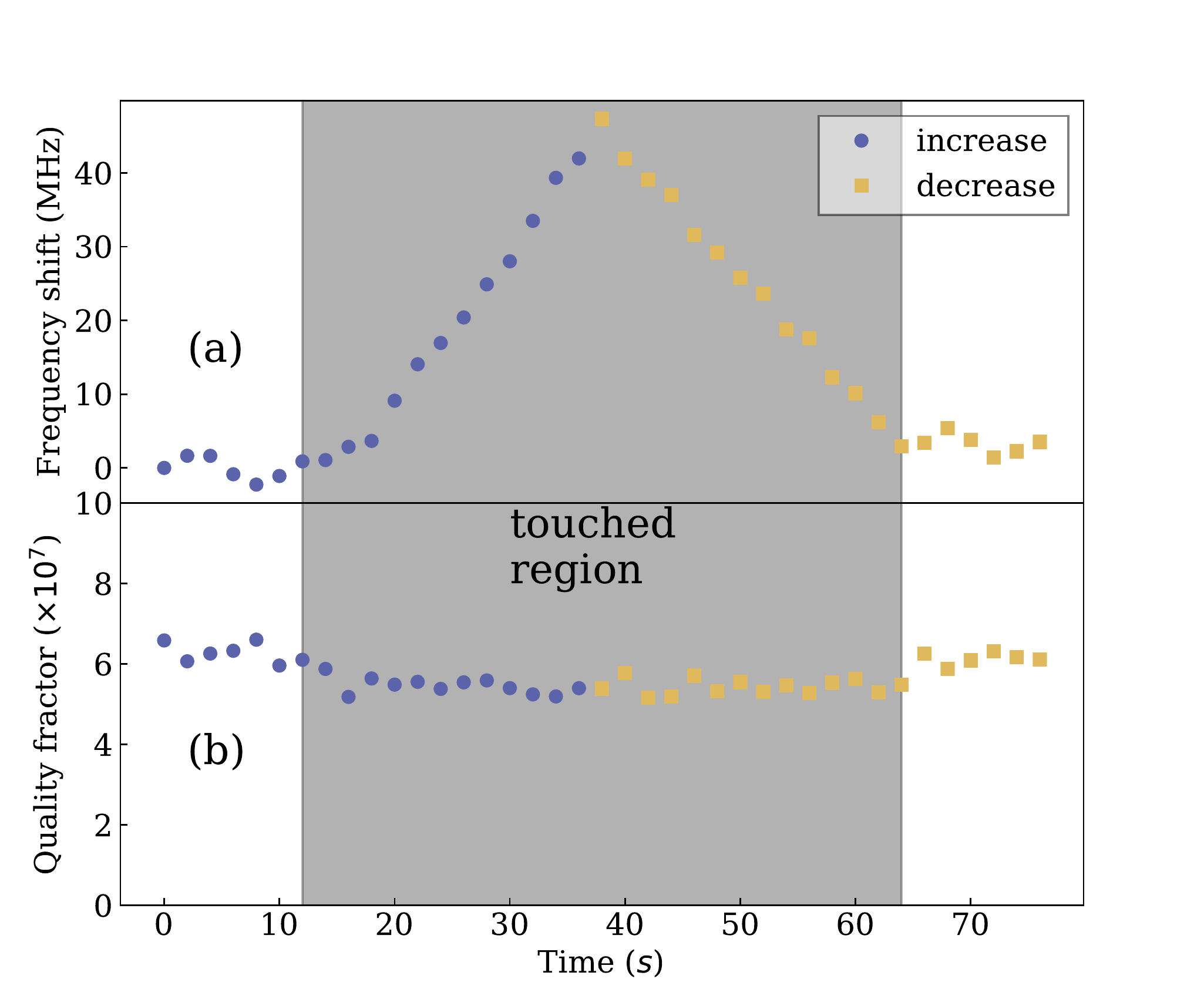}
\caption{(a) The relative frequency shift induced by the compression of nano-positioner. The working voltage was first increased to $V_m=$ \SI{3.8}{\V} with a step of $V_s=$ \SI{0.2}{\V} per two seconds and then decreased to \SI{0}{\V}. The nano-positioner touched the microsphere in the grey region. (b) The quality factor change before and after applying compression on the sphere. The Q drops \SI{12.7}{\percent} from its original of \SI{0.63d8}, without re-coupling the taper fiber. The blue circles and yellow square dots represent the increasing and decreasing the voltage, respectively. }
\label{fig:Q}
\end{figure}

The radius of the silica fused microsphere was \SI{103.50}{\um}. The nano-positioner was placed in front of the microsphere without touching it. By increasing the working voltage, the nano-positioner would move closer to the microsphere and touch it finally. In this measurement, the working voltage was increased from \SI{0}{\V} to maximum value $V_m=$ \SI{3.8}{\V} with a step of $V_s=$ \SI{0.2}{\V} per two seconds and then decreased to \SI{0}{\V} with the same speed. As shown in Fig. \ref{fig:Q}, we measured both the quality factor change and the frequency shift. The working voltage was \SI{1.2}{V} at \SI{12}{\second} which is high enough for the nano-positioner to touch the microsphere. With the nano-positioner moving further, the microsphere will be squeezed with nano-meter scale and the resonance will be tuned correspondingly [see gray part in Fig. \ref{fig:Q}]. In our experiment, the microsphere was fixed on a stage and stuck out about \SI{1}{\mm}. After the nano-positioner touched the microsphere, the direction of the fiber would slightly deviate from its original position, which would make the taper under-coupled or over-coupled with the microsphere. Since the mode volume of the WGM was constrained in the equator of the microsphere, which was far away from the touch point, the main reason for the drop of the Q would be the offset of the taper direction. The quality factor would have a \SI{12.7}{\percent} drop from \SI{0.63d8} to \SI{0.55d8}, as shown in Fig. \ref{fig:Q}.

\begin{figure}
\centering
\includegraphics[width=\linewidth]{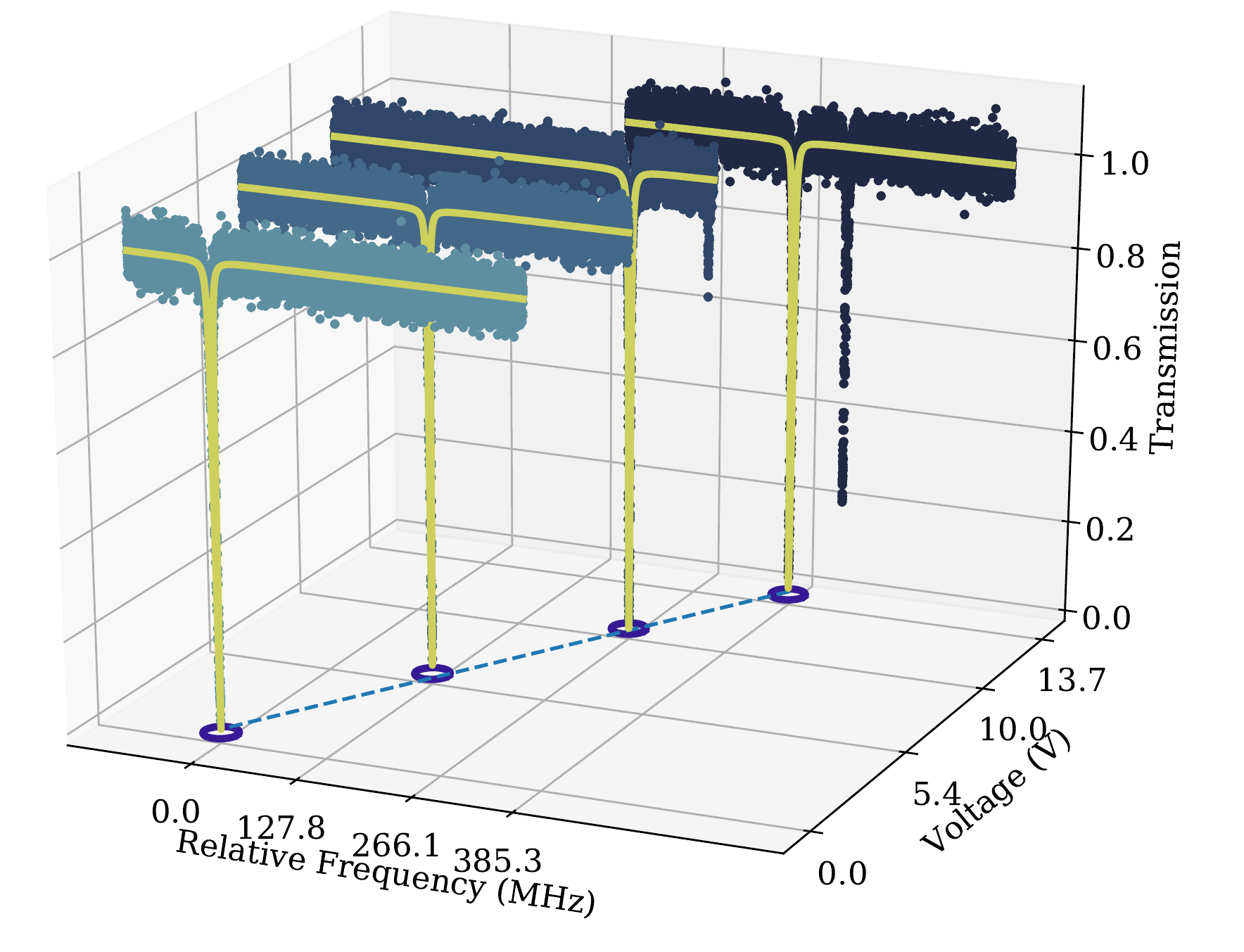}
\caption{The resonance shifts of the microsphere. For a microsphere with $R=$ \SI{103.50}{\um}, the frequency of WGM can be tuned to \SI{385.3}{\MHz} under a working voltage of $V_m=$ \SI{13.7}{\V}. By recoupling the taper fiber with the sphere on each measurement, we can keep the quality factor almost constant. The Q are (0.5, 0.61, 0.49 and 0.56) $\times 10^8$, with the increasing of voltage.}
\label{fig:all_range}
\end{figure}

In this experiment, we used another nano-positioner to adjust the position of the tapered fiber, which can recouple the taper with the microsphere without changing the resonance of the sphere. With the help of this extra nano-positioner, we can measure the maximum frequency shift of the microsphere. In Fig. \ref{fig:all_range}, we kept increasing the working voltage range from \SI{0}{\V} to \SI{13.7}{\V}, and achieved a \SI{385.3}{\MHz} shift compared to its original resonance frequency. By adjusting the tapered fiber while squeezing the microsphere, the Q could keep almost constant, which were (0.5, 0.61, 0.49 and 0.56) $\times 10^8 $ for the four working voltages that we measured [see in Fig. \ref{fig:all_range}(b)].

\section{Discussion and Conclusion}
In this strain tuning scheme, the frequency shift is induced by the geometry and the refractive index change of the microsphere cavity. The relation between them can be approximately described as \begin{equation}
 \frac{\Delta \nu}{\nu}=-\frac{\Delta R}{R}-\frac{\Delta n}{n}
\label{eq:relation}.
\end{equation} As shown in previous study, the effect of geometry change $\Delta R/R$ is much bigger than the index change \cite{har_mec}, so we can ignore the frequency shift induced by the index change. Because of the high precision nano-positioner, the deformation of the microsphere along the axial direction is around nano-meter, much less than the size of the sphere. The change of the equatorial semiaxes $\Delta R$ can be approximated as $\Delta R=\mu \Delta R_a$, where $\mu$ is the Poission ratio of silica and $\Delta R_a$ is the change of the axial semiaxes $R_a$. Taking red tuning for example, the frequency shift per step can be written as
\begin{equation}
	\Delta \nu=\frac{\nu\mu\Delta R_a}{R}=\frac{\nu\mu\alpha V_s}{R}
\label{eq:relation2},
\end{equation} where $\alpha=\Delta R_a/V_s$ is the piezoelectric strain constant of the nano-positioner. Here we can obtain that $\Delta \nu\propto V_s/R$, so it is straightforward to improve the frequency tuning resolution by adopting a bigger microsphere or using a higher precision nano-positioner. In this paper, we obtained an ultrahigh resolution frequency tuning with $\Delta \nu=$ \SI{650}{\kHz}, which is less than a half of the line width of the WGM that we used in this experiment.

\begin{figure}[htbp]
\centering
\includegraphics[width=\linewidth]{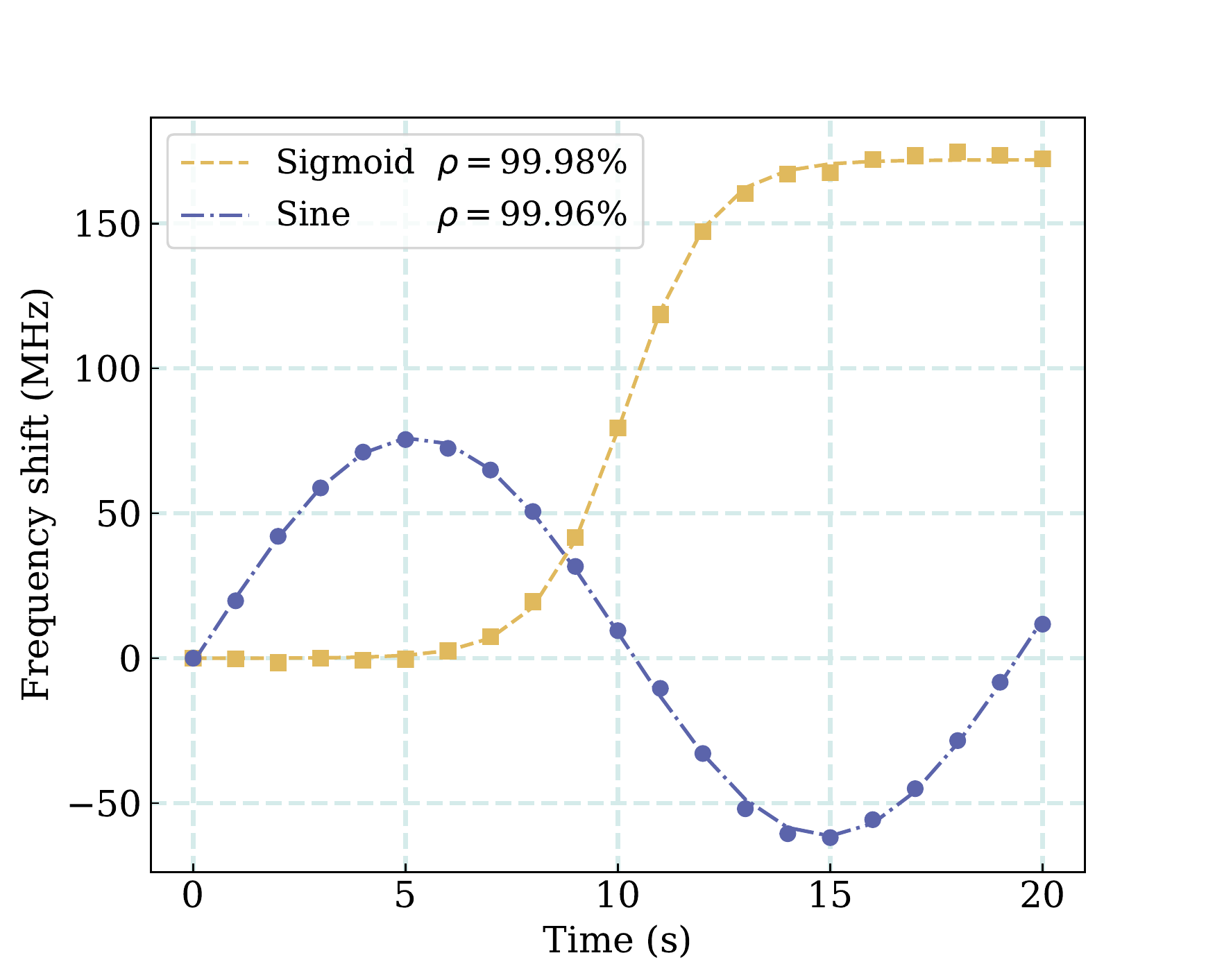}
\caption{Tuning the resonance frequency of the microsphere as a sine and a sigmoid function. The dash lines are the fitting result of these two functions. The Pearson correlation coefficients $\rho$ are \SI{99.98}{\percent} and \SI{99.96}{\percent} for sigmoid and sine function respectively.}
\label{fig:sin_sig}
\end{figure}

Up to now, all the working voltage that we adopt is changed linearly and the resonance frequencies are also tuned linearly. However, it is important and useful to tune the frequency nonlinearly. Combined with the high precise periodic tuning, we can easily achieve frequency time-periodic modulation, which is the key point in time-floquet system. Here we control the resonance tuning as a sine function. Because of the linear relationship between the working voltage of the nano-positioner and the frequency shift of the microsphere, we just need to add a sine signal voltage to the nano-positioner. According to Eq. \ref{eq:relation2}, the resonance of WGM can be written as
\begin{equation}
	\Delta \nu=\frac{\nu\mu\alpha \mathcal{F}(V)}{R}
\label{eq:relation3},
\end{equation} where $\mathcal{F}$ is an arbitrary continuous function, and $V$ is the working voltage. Furthermore, we tuned the resonance as a sigmoid function $S$, a vital activation function in neural network \cite{sigmoid} of machine learning, due to the linear relationship in Eq. \ref{eq:relation2}.

We represent in Fig. \ref{fig:sin_sig} the tuning of the WGM resonance frequency shift with the evolution of time in the first period. The two different kinds of frequency shifts here are nearly perfectly matched with the corresponding function shapes, with Pearson correlation coefficient $\rho$ of \SI{99.98}{\percent} and \SI{99.96}{\percent} respectively. By adjusting the amplitude of the working voltage, the amplitude of the frequency shift can also be easily tuned to anywhere within the tolerance. With the help of such a high precise functional control, the frequency can be tuned as any continuous wave. Because of the linear relationship between the frequency shift and the nano-positioner working voltage in Eq. \ref{eq:relation2} (also shown in Fig. \ref{fig:u_d_p}), the excellent periodically frequency tuning as well as a sine or a sigmoid function tuning could be achieved.

In this paper, \SI{650}{\kHz} resolution of WGM resonance tuning has been achieved, much better than previous works. Furthermore, a sine wave and sigmoid function resonance tuning with 99\% accuracy have also been successfully accomplished. This ultrahigh resolution and ultrahigh precise AFRETOM make it much easier to do frequency matching of WGM with ultrahigh quality factor and also pave the way for time-floquet tuning experiment in cavity QED experiments.

This work was supported by the National Natural Science Foundation of China (20171311628), National Key Research and Development Program of China (2017YFA0303700) and Beijing Advanced Innovation Center for Future Chip (ICFC).


\begin{thebibliography}{99}
\bibitem{vahala_mc} K. J. Vahala, nature 424, 839 (2003).
\bibitem{yl_sensing} J. Zhu, S. K. Ozdemir, Y.-F. Xiao, L. Li, L. He, D.-R. Chen, and L. Yang, Nat. Photonics 4, 46 (2009).
\bibitem{xyf_sensing} J. M. Ward, Y. Yang, F. Lei, X.-C. Yu, Y.-F. Xiao, and S. N. Chormaic, Optica 5, 674 (2018).
\bibitem{sqh_sensing} N. Zhang, Z. Gu, S. Liu, Y. Wang, S. Wang, Z. Duan, W. Sun, Y.-F. Xiao, S. Xiao, and Q. Song, Optica 4, 1151 (2017).
\bibitem{xyf_sensing2} X.-C. Yu, Y. Zhi, S.-J. Tang, B.-B. Li, Q. Gong, C.-W. Qiu, and Y.-F. Xiao, Light. Sci. \& Appl. 7, 18003 (2018).
\bibitem{yl_sensing2} X. Xu, W. Chen, G. Zhao, Y. Li, C. Lu, and L. Yang, Light. Sci. \& Appl. 7, 62 (2018).
\bibitem{mk_qed} H. Mabuchi and H. J. Kimble, Opt. Lett. 19, 749 (1994).
\bibitem{vahala_qed} D. Alton, N. Stern, T. Aoki, H. Lee, E. Ostby, K. Vahala, and H. Kimble, Nat. Phys. 7, 159 (2011).
\bibitem{hailin_qed} M. Larsson, K. N. Dinyari, and H. Wang, Nano letters 9, 1447 (2009).
\bibitem{bf_nonlinear} F. Bo, J. Wang, J. Cui, S. K. Ozdemir, Y. F. Kong, G. Q. Zhang, J. J. Xu, and L. Yang, Adv. Mater. 27, 8075 (2015).
\bibitem{xyf_nonlinear} Q. T. Cao, H. M. Wang, C. H. Dong, H. Jing, R. S. Liu, X. Chen, L. Ge, Q. H. Gong, and Y. F. Xiao, Phys. Rev. Lett. 118 (2017).
\bibitem{hailin_opto} Y. S. Park and H. L. Wang, Nat. Phys. 5, 489 (2009).
\bibitem{kipp_opto} S. Weis, R. Riviere, S. Deleglise, E. Gavartin, O. Arcizet, A. Schliesser, and T. J. Kippenberg, Science. 330, 1520 (2010).
\bibitem{wm_opto2} X. F. Jiang, M. Wang, M. C. Kuzyk, T. Oo, G. L. Long, and H. L. Wang, Opt. Express 23, 27260 (2015).
\bibitem{vahala_opto} T. Carmon, H. Rokhsari, L. Yang, T. J. Kippenberg, and K. J. Vahala, Phys. Rev. Lett. 94 (2005).
\bibitem{wm_opto} M. Wang, Y.-Z. Wang, X.-S. Xu, Y.-Q. Hu, and G.-L. Long, Opt. Express 27, 63 (2019).
\bibitem{cy_ln} M. Wang, R. Wu, J. Lin, J. Zhang, Z. Fang, Z. Chai, and Y. Cheng, Quantum Eng. (2019). Https://doi.org/10.1002/que2.9.
\bibitem{or_ln} R. Osellame, Quantum Eng. (2019). Https://doi.org/10.1002/que2.11.
\bibitem{dch_opto} Z. Shen, Y. L. Zhang, Y. Chen, C. L. Zou, Y. F. Xiao, X. B. Zou, F. W. Sun, G. C. Guo, and C. H. Dong, Nat. Photonics 10, 657 (2016).
\bibitem{hailin_deformed} Y. S. Park and H. L. Wang, Opt. Express 15, 16471 (2007).
\bibitem{frank_biosensing} F. Vollmer and S. Arnold, Nat. Methods 5, 591 (2008).
\bibitem{wang_scibut} T. Wang, X.-F. Liu, Y. Hu, G. Qin, D. Ruan, and G.-L. Long, Sci. bulletin 63, 287 (2018).
\bibitem{wc_microring} Y.-P. Gao, T.-J. Wang, C. Cao, and C. Wang, Photon. Res. 5, 113 (2017).
\bibitem{sumetsky_mb} M. Sumetsky, Y. Dulashko, and R. S. Windeler, Opt. Lett. 35, 1866 (2010).
\bibitem{yy_mb} Y. Yang, S. Saurabh, J. Ward, and S. N. Chormaic, Opt. Lett. 40, 1834 (2015).
\bibitem{lxf_chiral} X. F. Liu, T. J. Wang, Y. P. Gao, C. Cao, and C. Wang, Phys. Rev. A 98 (2018).
\bibitem{yl_eit} B. Peng, S¸ . K. Özdemir, W. Chen, F. Nori, and L. Yang, Nat. Commun. 5, 5082 (2014).
\bibitem{fxd_eit} S. Zhu, L. Shi, S. X. Yuan, R. L. Ma, X. L. Zhang, and X. D. Fan, Nanophotonics. 7, 1669 (2018).
\bibitem{ktt_tf} T. T. Koutserimpas and R. Fleury, Phys. Rev. Lett. 120 (2018).
\bibitem{hailin_tempe} Y. S. Park, A. K. Cook, and H. L. Wang, Nano Lett. 6, 2075 (2006).
\bibitem{wt_eit} T. Wang, Y.-Q. Hu, C.-G. Du, and G.-L. Long, Opt. Express 27, 7344 (2019).
\bibitem{oliver_gas2} R. Henze, T. Seifert, J. Ward, and O. Benson, Opt. Lett. 36, 4536 (2011).
\bibitem{oliver_gas} R. Henze, J. M. Ward, and O. Benson, Opt. Express 21, 675 (2013).
\bibitem{hailin_mec} K. N. Dinyari, R. J. Barbour, D. A. Golter, and H. L. Wang, Opt. Express 19, 17966 (2011).
\bibitem{dch_mec} Z. H. Zhou, C. L. Zou, Y. Chen, Z. Shen, G. C. Guo, and C. H. Dong, Opt. Express 25, 4046 (2017).
\bibitem{wolf_mec} W. von Klitzing, R. Long, V. S. Ilchenko, J. Hare, and V. Lefevre-Seguin, Opt. Lett. 26, 166 (2001).
\bibitem{har_mec} V. S. Ilchenko, P. S. Volikov, V. L. Velichansky, F. Treussart, V. Lefevre-Seguin, J. M. Raimond, and S. Haroche, Opt. Commun. 145, 86 (1998).
\bibitem{oskar_conv} J. T. Hill, A. H. Safavi-Naeini, J. Chan, and O. Painter, Nat. Commun. 3, 1196 EP (2012).
\bibitem{Zhang_shortcuts} H. Zhang, X.-K. Song, Q. Ai, H. Wang, G.-J. Yang, and F.-G. Deng, Opt. Express 27, 7384 (2019).
\bibitem{dch_conv} C. Dong, V. Fiore, M. C. Kuzyk, L. Tian, and H. Wang, Annalen der Physik 527, 100 (2015).
\bibitem{xf_therm} X. F. Liu, F. C. Lei, M. Gao, X. Yang, G. Q. Qin, and G. L. Long, Opt. Lett. 41, 3603 (2016).
\bibitem{sigmoid} J. Han and C. Moraga, “The influence of the sigmoid function parameters on the speed of backpropagation learning,” in International Workshop on Artificial Neural Networks, (Springer), pp. 195–201.

\end{thebibliography}
\end{document}